\documentclass[aps,pre,twocolumn,showpacs,floatfix,12pt]{revtex4}
\usepackage{graphicx}
\begin{document}


%
%

\title{Critical behavior in an evolutionary model of altruistic behavior with
social structure}
\author{V\'{\i}ctor M. Egu\'{\i}luz} \email{victor@imedea.uib.es}
\homepage{http://www.imedea.uib.es/~victor} \affiliation{Instituto
Mediterr\'aneo de Estudios Avanzados IMEDEA (CSIC-UIB), E-07122 Palma de
Mallorca, Spain} \author{Claudio J. Tessone} \email{tessonec@ethz.ch}
\homepage{http://www.imedea.uib.es/~tessonec} \affiliation{Chair of Systems Design, ETH Z\"urich,
Kreuzplatz 5, CH-8032 Z\"urich, Switzerland} \affiliation{Instituto
Mediterr\'aneo de Estudios Avanzados IMEDEA (CSIC-UIB), E-07122 Palma de
Mallorca, Spain}

\date{\today}

%
\begin{abstract} Experimental studies have shown the ubiquity of altruistic
behavior in human societies. The social structure is a fundamental
ingredient to understand the degree of altruism displayed by the
members of a society, in contrast to individual-based features,
like for example age or gender, which have been shown not to be
relevant to determine the level of altruistic behavior. We explore
an evolutionary model aiming to delve how altruistic behavior is
affected by social structure. We investigate the dynamics of
interacting individuals playing the Ultimatum Game with their
neighbors given by a social network of interaction. We show that a
population self-organizes in a critical state where the degree of
altruism depends on the topology characterizing the social
structure. In general, individuals offering large shares but in
turn accepting large shares, are removed from the population. In
heterogeneous social networks, individuals offering intermediate
shares are strongly selected in contrast to random homogeneous
networks where a broad range of offers, below a critical one, is
similarly present in the population.
\end{abstract}

\keywords{Altruistic behavior; Social Networks; Self-organization}

\maketitle

\section{Introduction}

How cooperative behavior emerges among interacting individuals is
a long-standing problem that has attracted, starting from Darwin,
the attention of a large number of researches
\cite{Darwin71,Gould02,Pennisi05}. In the context of Game Theory,
different mechanisms and models have been proposed to explain the
observed cooperative behavior, two prominent examples being the
Ultimatum Game \cite{Guth90} and the Prisoner's Dilemma
\cite{Axelrod1984}. Theoretical studies, have shown that,
selection at the individual level, may lead to altruistic behavior
\cite{Sanchez04}, in contrast to the  general belief that only
group selection can give raise to altruism.

Aiming at understanding the mechanisms leading to altruism as the
core of cooperative behavior, the Ultimatum Game is one of the
paradigmatic theoretical games used to understand their
interrelation. The simplicity of this game has allowed to obtain a
large set of experimental results that clearly show the presence
of altruistic behavior, in the form of {\em altruistic
punishment}. Altruistic punishment, meaning that individuals react
to an unfavorable action by an opponent although the punishment is
costly for them and yields no material gain, is a key ingredient
for the explanation of cooperation as it emerges if altruistic
punishment is possible, while breaks down if it is ruled out
\cite{Fehr02}. The ultimatum game consists of two agents, who have
to share a given amount of money. One of them, the {\em proposer},
makes an offer on how to share the money to the other agent, the
{\em responder}. The proposer can only make one offer. The
responder decides whether he accepts or rejects the offer. If the
responder accepts, the money is shared as proposed; otherwise,
none of them get anything. Given that a narrowly rational
responder would accept no matter what he has been offered
--something is better than nothing--, a narrowly rational proposer
would offer to the responder the minimum amount. However, real
agents behave differently: offers are typically close to a 50-50
ratio and offers below 20\% are typically rejected
\cite{JHetal:2005,JHetal:2006,EE:2004}.

Recently, an extensive research performed among small societies
around the world \cite{HBB01} has shown that social structure is a
key element in determining the degree of cooperation among its
members \cite{JHetal:2005,JHetal:2006}. On the one hand,
individual-based features seem not to be relevant in order to
determine the degree of cooperation. On the other hand, in the
Ultimatum Game, the proposals and rejection levels are different
depending on the social structure. For example, societies based on
cooperation and sharing of food, show higher offering levels. From
the theoretical point of view the effect of structured populations
in social dilemmas, e.g., the prisoner's dilemma, public good
games, or snowdrift games, has been analyzed mainly from the
perspective of spatially extended populations
\cite{Hauert06a,Hauert06b,Nowak92,Page00b}. In the simplest case,
local interactions are considered in regular lattices where each
individual interacts only within its local neighborhood in
contrast to global random interactions considered in well mixed
populations. However, recent progress in this area has shown that
many social and biological interaction networks are not regular,
as typically used in theoretical models, but display the
small-world behavior and broad degree distributions
\cite{Albert02,Amaral00,Hamilton:2007,Onnela:2007,Palla:2007,Szabo07}.

The question we address here is precisely how social structure
 affects the degree of altruistic behavior. This paper is organized
as follows: In the next Section we define the evolutionary
ultimatum game model in a complex network. In
Section~\ref{single-scale}, following the tradition of spatial
games, we first consider interactions given by a regular lattice;
later extending our analysis to random and small-world networks,
which are characterized by single-scale degree distributions; in
Section~\ref{scale-free}, we also consider scale-free networks.
Finally we discuss our results and draw the conclusions in
Section~\ref{conclusion}.

\begin{figure}
\centerline{\includegraphics[width=.4\textwidth,angle=-90]{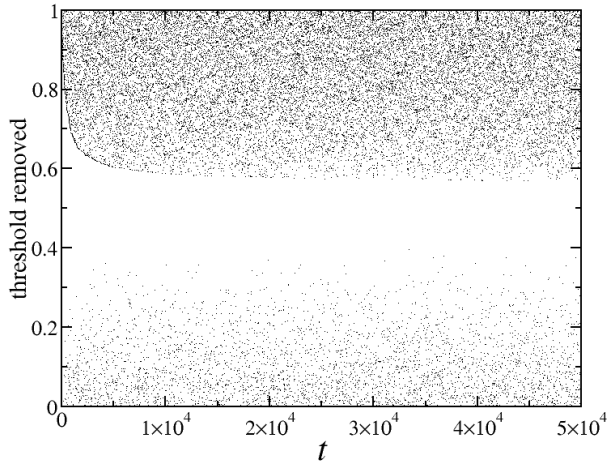}}
\caption{For each time step, a data point indicates the threshold of the agent
with the lowest payoff. The population is composed of $10^4$ agents who
interact in a one-dimensional lattice.}
\label{fig:time}
\end{figure}

\section{The model}

A set of $N$ agents are arranged in the nodes of a network. We set
1 unit the amount to be shared in each interaction. Each agent $i$
is characterized by a threshold $T_i \in [0,1]$: as responder it
indicates the minimum amount he will accept; as proposer, it also
defines the amount of money he will offer. This situation is
usually named as {\em empathy} \cite{Preston02}, and has been
shown to enhance fairness in some situations in the ultimatum game
\cite{Page00a,Page02,Sanchez04}. Based on experimental evidence
reported previously, we will assume that a fair offer-acceptance
(threshold values around 50\%) represents altruistic behavior.

The model runs as follows: at each time step all agents play with
all their neighbors synchronously. Thus for each interaction link
between two neighboring agents $(i,j)$,
\begin{itemize}
 \item If the offer $T_i$ is above the threshold of agent $j$, $T_i > T_{j}$,
 then the offer $T_i$ is accepted: agent $j$ increases his payoff by $T_i$ while
 agent $i$'s payoff increases by $1-T_i$. The payoff obtained by agent $i$ ($j$)
 from the interaction with his neighbor $j$ ($i$) is $\Pi_{ij}= 1-T_i$
 ($\Pi_{ji}= T_i$);
 \item Otherwise if the offer $T_j$ is above the threshold of agent $i$, $T_{j}
 > T_{i}$, then the offer $T_j$ is accepted: agent $i$'s payoff increases by $T_{j}$
 while agent $j$'s payoff increases by $1-T_{j}$. The payoff obtained by agent
 $i$ ($j$) from the interaction with his neighbor $j$ ($i$) is $\Pi_{ij}= T_j$
 ($\Pi_{ji}= 1- T_j$).
\end{itemize}

The payoff obtained by agent $i$ after interaction with the
neighbors $j$ in his neighborhood ${\cal V}(i)$ is thus $\Pi_i =
\sum_{j\in {\cal V}(i)} \Pi_{ij}$. In the unlikely event that two
neighboring agents thresholds are the same, one of the offers is
selected and accepted at random. After each round, a selection
rule is applied to the system: the agent with the lowest payoff in
the population and its immediate neighbors, determined by the
network, are replaced by new agents with randomly chosen
thresholds \cite{bak93}. We then let the system evolve resetting
the payoffs of all agents to zero.

An alternative description of the model could set the interaction
between two agents as a single event in which agent $i$ acts as
proposer and $j$ as responder. However as long as the update is
synchronous and every agent plays as proposer and responder with
all its neighbors, this description and the dynamic rules (i)-(ii)
are equivalent.

\begin{figure}
\centerline{\includegraphics[width=.4\textwidth,angle=-90]{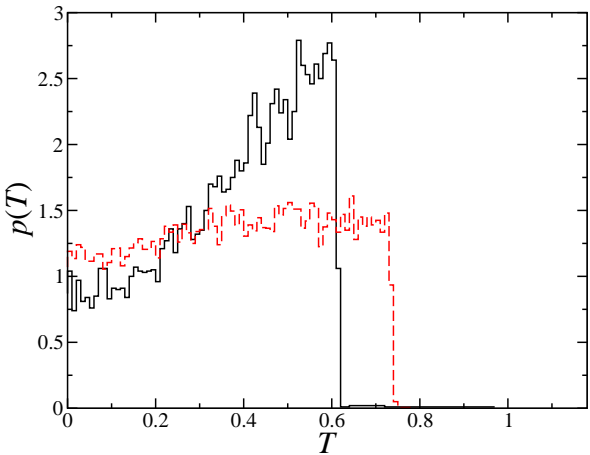}}
\caption{Distribution of thresholds for $10^4$ agents in a
one-dimensional lattice (continuous line) and in a random network (dashed line)
after $4 \times 10^6$ iterations. The dynamics selects threshold values below
the critical threshold $T_c =0.56$ in the one-dimensional lattice and $T_c=0.88$ in the random network.
}
\label{fig:threshold}
\end{figure}

\section{Results}
\label{single-scale}

\subsection{Single-scale interaction networks}
Following previous studies of spatial games
\cite{Nowak92,KD:1998,Santos06}, we first consider a
one-dimensional lattice where each agent interacts with his two
nearest neighbors. We have run simulations for populations in the
range $N=10^3$ to $10^4$ agents, finding consistent results. The
initial thresholds are randomly selected from a uniform
distribution in the range $[0,1]$. In Fig.~\ref{fig:time} we
display the threshold of the agent with the lowest payoff in the
population at each time step. Agents with high and low thresholds
are removed from the population. On the one hand, agents with high
thresholds make large offers that are likely to be accepted by
their neighbors, contributing to the neighbors payoff. However
quite unlikely they receive large enough offers to be accepted due
to their high threshold. This behavior can be illustrated
calculating the expected payoff per interaction in a completely
mixed population,
\begin{equation}
\langle \Pi(T)\rangle = T(1-T) + \frac{1-T^2}{2}~. \label{eq1}
\end{equation}
In the limiting case of agents with threshold $T=1$, they obtain
on average a payoff close to 0. On the other hand, the opposite
situation is observed for agents with low threshold: their offers
are hardly accepted while they accept most of the offers they
receive. In the limiting case of agents with threshold $T=0$, they
obtain on average a payoff close to 1/2 per neighbor. Thus agents
with low thresholds have in average a larger payoff than agents
with high thresholds and are able to survive. Although the
previous argument has been obtained for a completely mixed
population, we observe that it is still valid in the regular case
(Fig.~\ref{fig:threshold}). Thus, after a transient, the
distribution of thresholds reaches a stationary distribution.
Thresholds above a critical value are removed from the population.
For the one-dimensional lattice considered here, a critical
threshold value is obtained $T_c = 0.56\pm 0.01$. Below this
critical value the distribution of thresholds is not uniform: it
increases approaching the critical value. The payoff distribution
also shows a nonuniform distribution: only values above a critical
value $\Pi_c=1.76$ are found, displaying a maximum at a value
around $\Pi \simeq 2$.

\begin{figure}[t]
\centerline{\includegraphics[width=.4\textwidth,angle=-90]{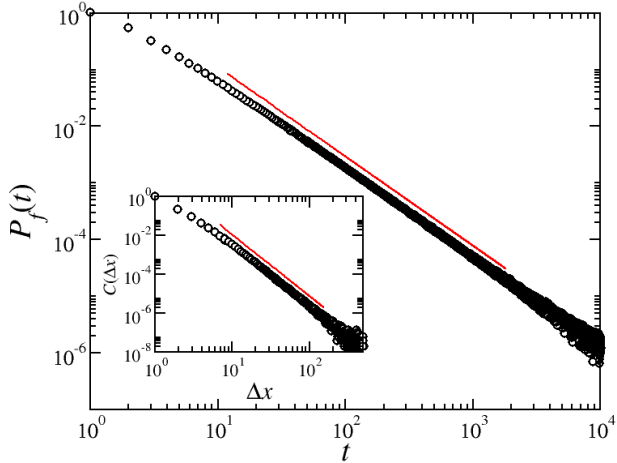}}
\caption{First return time distribution for a system of $10^4$ agents in a
one-dimensional lattice. The solid line is a power-law fit with an exponent
$\alpha=1.57$. Inset: distribution of the distance $\Delta x$ between two
agents getting the lowest payoff consecutively.
The solid line is a power-law fit with an exponent $\gamma=3.16$.}
\label{fig:firstr}
\end{figure}

The question we next address is whether the ultimatum model
self-organizes in a critical state. In order to characterize the
dynamics, we have measured the distribution of the distance
$\Delta x$ between two consecutive selection events, $C(\Delta
x)$, and the first return time distribution, $P_f(t)$, the time
elapsed between two selection events affecting the same agent. The
results are plotted in Fig.~\ref{fig:firstr}. In both cases the
tails of the distributions are well fitted by power-laws. For the
spatial correlation
\begin{equation}
 C(\Delta x) \sim \Delta x^{-\gamma}~,
\end{equation}
 with $\gamma=3.16 \pm 0.1$, and for the first return time
\begin{equation}
P_f(t) \sim t^{-\tau}~,
\end{equation}
with $\tau=1.57$.

\begin{figure}
\centerline{\includegraphics[width=.4\textwidth,angle=-90]{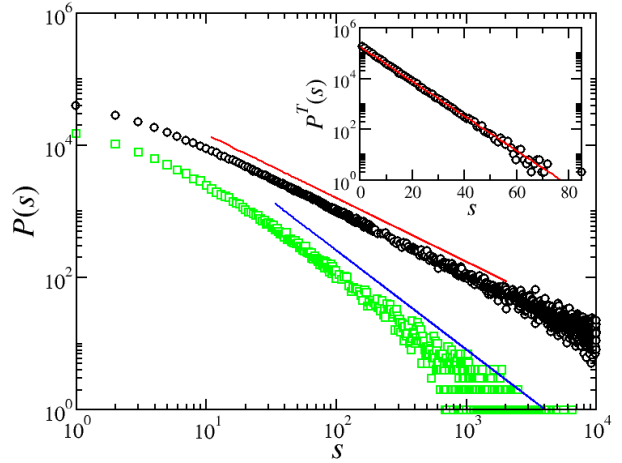}}
\caption{Avalanche size distribution $P(s)$ a system of $10^4$ agents in a
one-dimensional lattice (circles) and in small-world networks (squares) with a rewiring
probability $p=0.15$. The solid lines is a power-law fit with an exponent
$\alpha=1.0$ and $\alpha=1.5$, respectively. Inset: the
avalanche size distribution $P^{T}(s)$ considering the thresholds instead
of payoffs for the definition of an avalanche in a one-dimensional lattice.
The solid line corresponds to the best exponential fit.
}
\label{fig:ava}
\end{figure}

These results suggest that the system self-organizes in a critical
state where the distribution of avalanches is also a power-law.
The critical state would emerge despite the nonuniform
distribution of thresholds (and payoffs). An avalanche is
typically defined as follows: it starts when the lowest payoff
gets larger than a preset value $\Pi^*$ (close to the critical
payoff), and stops when it drops below this value. The size, $s$,
of an avalanche is the number of time steps it lasts. In
Fig.~\ref{fig:ava}, we show the distribution of avalanche sizes
when we use a payoff $\Pi^* = 1.76$ as the indication of an
avalanche. The probability distribution displays a power-law decay
\begin{equation}
P(s) \sim s^{-\alpha},
\end{equation}
with an exponent $\alpha = 1.0$.

It is worth noting that if we had used the threshold value, $T$,
as an indication of when an avalanche starts and stops, the
distribution of avalanches would have decayed exponentially (see
inset of Fig.~\ref{fig:ava}). This behavior reflects that the
thresholds being removed are not always above a critical value (as
shown in Fig.~\ref{fig:time}): agents with low thresholds are
removed often from the population. Thus, the fitness of an agent
is given by its payoff.

\begin{figure}[t]
\centerline{\includegraphics[width=.4\textwidth]{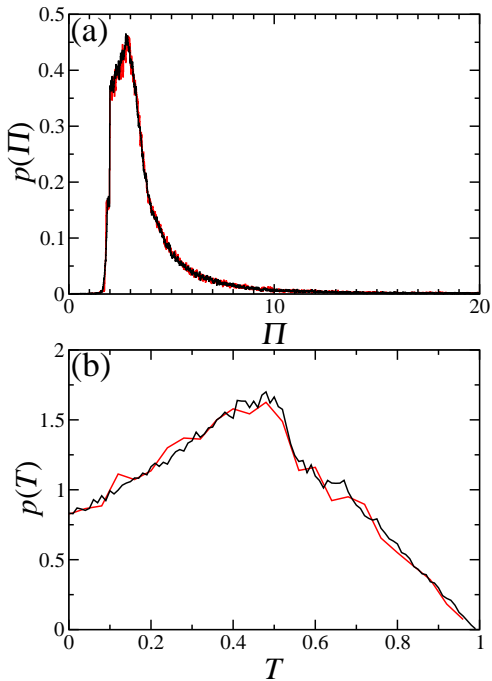}}
\caption{(a) Distribution of payoffs
and (b) distribution of thresholds for $10^4$ agents in a scale-free network
in the asymptotic state after $10^6$ iterations.
The distributions are averaged over 100 realizations.}
\label{fig:U:pt}
\end{figure}

Regular lattices are just a crude simplification of social and
biological interaction. More realistic models of interaction
networks include the small-world behavior: the average distance
between agents in the network is similar to the one obtained in a
random network, and the clustering, the fraction of neighbors of
an agent that are also neighbors between them is large, as occurs
in a regular lattice. We have performed simulations in small-world
networks generated by rewiring the links of a one-dimensional
lattice, using the algorithm introduced in Ref.~\cite{Maslov03} in
order to keep all the agents with the same number of links: With a
probability $p$, two edges exchange their end nodes. The
distribution of thresholds and payoff are similar to the
one-dimensional lattice, as shown in Fig.~\ref{fig:threshold}. It
can be seen that for the limiting case of a random network, the
distribution of thresholds is broader, and has a higher critical
threshold. In this case, the system also self-organizes in a
critical state where the avalanche size distribution displays a
power-law scaling. Similarly to the one-dimensional lattice, the
avalanche size distribution (when considering the threshold as the
dynamical variable) is not power-law, but exponential. For the
small-world networks, the distribution of avalanche size is also
power-law with an exponent that depends on the rewiring
probability $p$. For instance, for a system of $10,000$ agents we
find an exponent $\alpha =1.5$, for a value $p=0.15$
(Fig.~\ref{fig:ava}).

\begin{figure}[t]
\centerline{\includegraphics[width=.4\textwidth,angle=-90]{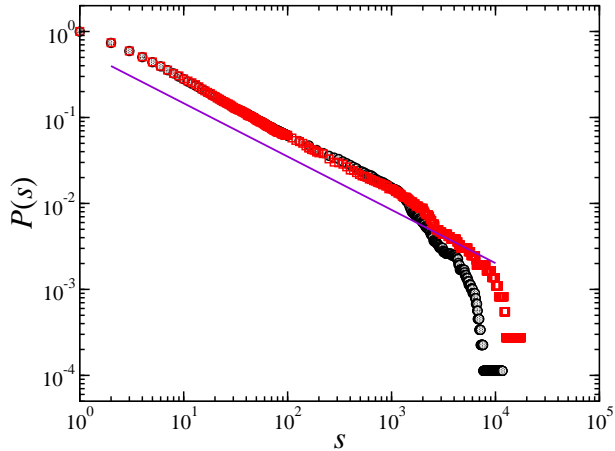}}
\caption{The cumulative distribution of avalanches in a scale-free network
with $10^4$ (circles) and $2\times10^4$ agents. The distribution is well
fitted by a power-law with an
exponential cutoff. The solid line corresponds to a power-law fit,
leading to an exponent $\alpha=1.62$.
The preset payoff that defines when an avalanche starts and ends
is $\Pi^*=1.77$. Averages are obtained over 100 realizations.}
\label{fig:U:av}
\end{figure}

\subsection{Scale-free interaction networks}
\label{scale-free}

In all the results presented so far, the network of interactions
is such that the number of links of each node, its {\em degree},
is constant for all the agents. Beyond the small-world behavior,
another important ingredient of interaction networks is that they
often display a scale-free degree distribution. We now turn into
the study of the results of this model when the topology of
interactions is not a regular one, but a scale-free one. As the
model for the generation of the network, we used the
Barab\'asi-Albert algorithm \cite{Barabasi99a,Barabasi99b},
generated as follows: starting from a fully connected network of
size $m$, at time $t$ a node is added, and attached to $m$
existing nodes, where the probability to be attached to a node is
proportional to its degree. This algorithm generates networks with
a power-law degree distribution with an exponent $\gamma = 3$. We
have fixed the value $m=2$. Once the network is grown, it is kept
fixed and the dynamics is played as indicated by rules (i) and
(ii).

In Fig.~\ref{fig:U:pt}(a) we plot the stationary distribution of
payoffs in the population. There is a well defined critical
payoff, below which the agents are removed. This critical value is
$\Pi_c=1.75 \pm 0.02$. For large payoffs, the distribution decays
as a power-law with an exponent of $3$, reflecting the decay of
the degree distribution of the network. For the distribution of
thresholds in the population (Fig.~\ref{fig:U:pt}(b)), there is
not cutoff in this distribution. All thresholds in the range
$[0,1]$ are present, with a maximum in the distribution around
$T\simeq 0.5$. The distribution is highly asymmetric: it
approaches zero for thresholds close to 1, while it reaches a
finite fraction in the limit $T\to 0$. Thus, agents with low
threshold values have more chances to survive. This is in
agreement with the analytical argument in Eq.~\ref{eq1}. In
scale-free networks, the system also exhibits a power-law
distribution of the size of the avalanches (Fig.~\ref{fig:U:av}).
In this case, the payoff we set to define an avalanche is
$\Pi^*=1.77$ obtaining an exponent $\tau=1.62$ for the a power-law
scaling.

\begin{figure}
\centerline{\includegraphics[width=.4\textwidth,angle=-90]{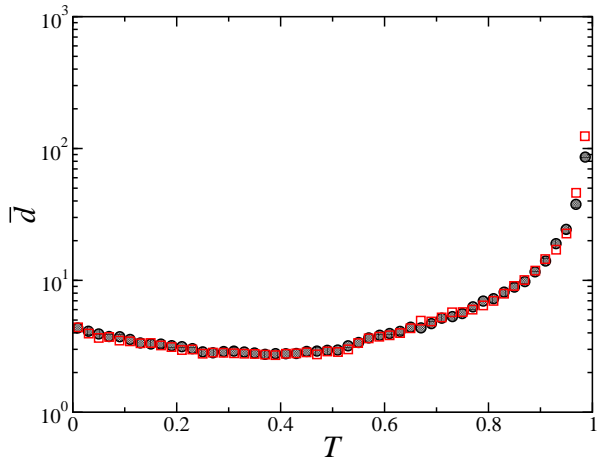}}
\caption{Average degree $d$ of agents having a given threshold
$T$ when $10^4$ agents interact in a scale-free network.
Averages are obtained over 50 realizations.}
\label{fig:U:tmed}
\end{figure}

Why can in the scale-free network agents with high thresholds
survive better than in a regular or random network? To address
this question we have analyzed the average degree of the agents
grouped according to their threshold, and the results are shown in
Fig.~\ref{fig:U:tmed}. It can be seen  that agents with high
threshold values are more likely to survive if they are located at
the hubs of the network. The reason for this is that they can
accumulate payoff via interaction with a large number of agents.

It is worth noting that when the network is highly heterogeneous,
for example when the degree distribution is scale-free, the
dynamics depends on whether the payoff of each agent is normalized
by its degree. For normalized payoffs in the scale-free networks
described previously the critical threshold depends on the number
of agents $N$. In the limit of large $N$ the critical threshold
tends to 1, in contrast to the system size independent critical
threshold reported previously, and the dynamics displays power-law
distribution of avalanches when the threshold used to define an
avalanche is chosen depending on $N$. A deeper analysis of the
normalized payoff case goes beyond the scope of this paper.

\section{Discussion and conclusions}
\label{conclusion}

How cooperation and altruism emerge among individuals is a
withstanding question that has attracted much attention in the
last years. Also the relevance of the complex social organization,
and the concomitant network of interactions in supporting
altruistic behavior is an open question
\cite{JHetal:2005,JHetal:2006}.

We have proposed an evolutionary ultimatum game with local
interactions that self-organizes in a critical state. To analyze
how social structure influences the degree of cooperation we have
considered different topologies for the network of interactions.
Assuming fairness as offer-acceptance around 50\%, the amount of
altruistic behavior in the Ultimatum Game is reflected in the
distribution of thresholds in the population. In regular and
single-scale networks, high thresholds leading to low payoffs are
removed, while intermediate thresholds are selected in the
populations. In scale-free networks, the distribution of
thresholds displays a maximum around a value of 50\%, decaying for
lower and larger threshold values. Comparing with experimental
results in small societies, both settings, the random and
scale-free topologies, capture  the experimental findings where
offers around 50\% are the most common. However, in the scale-free
networks the distribution of thresholds covers all the range of
threshold values decaying slowing from the maximum around 50\%, in
contrast to the sharp cutoff obtained in the random case. Thus, a
hierarchical social structure may explain better the patterns of
offer-acceptance found in some societies
\cite{Hamilton:2007,JHetal:2006}.

From a dynamical point of view, in all the different complex
networks we have analyzed, including regular lattices,
small-world, random and scale-free networks, the distribution of
avalanches displays a power-law scaling with an exponent that
depends on network topology. This feature is typically a signature
of self-organized criticality. Many complex systems in nature are
found to display this phenomenon \cite{Bak96,Jensen98}. They
characterize long-range correlations in a system, similar to the
behavior near a critical point in a phase transition. The model
introduced here can be compared with other evolutionary models.
The exponents that characterize the dynamics is the same as in the
simple model of evolution proposed by Bak and Sneppen (BS model)
\cite{bak93}. For the one-dimensional lattice, the exponents
characterizing the first return time, spatial correlation and
avalanche distribution for the ultimatum model introduced here are
the same as for the BS-model; for random networks, the exponent of
the avalanche size distributions corresponds to the mean-field
exponent of the BS model. However, there is a crucial difference
between the BS-model and the one introduced here: while in the
BS-model the fitness is directly assigned to the agents randomly
from a uniform distribution. In our model, the fitness is the
outcome of the interactions and the distribution of thresholds is
an emerging property of it.

The results presented here complement previous theoretical works
on the emergence of cooperation. Among the most used theoretical
games to study the emergence of cooperation in social sciences,
the Prisoner's Dilemma is perhaps the most paradigmatic example.
In this model, it has been shown that local interactions among
agents can lead to a cooperative behavior in the so-called {\em
spatial games} \cite{Nowak92}. In evolutionary models, it was also
found \cite{KD:1998} that self-organized criticality can be
present for this system. These results triggered the analysis of
spatial games in different network topologies and strategies of
the agents, aiming at uncovering the conditions under which
cooperation can arise \cite{Santos06}. In particular scale-free
networks have been shown to sustain cooperation \cite{KD:1998}.
Together with our results, hierarchical social structures,
represented for instance by scale-free networks of interaction,
suggest that primitive societies displaying this kind of
interactions favored the emergence of altruistic behavior. To
elucidate whether the interaction patterns facilitated altruism or
altruistic behavior led to complex interaction patterns we need to
incorporate more realistic ingredients to the models as for
instance the possibility of removal and establishment of social
ties depending on the outcome of the interaction \cite{ve:2005}.
Another open question is the evolution of empathy itself. In this
work we have assumed that empathy has evolved before the
thresholds are selected. However we expect that empathy will
co-evolve simultaneously with the thresholds. Other extensions of
the model could consider, for example, several repeated
interactions between agents before selection. In this case, it
could be argued that a repeated interaction scheme could allow for
an adaptation of the threshold by the least successful agents. The
interplay between adaptation and selection is an open question to
be addressed in future works.

In summary, experiments have shown that altruistic behavior is
common in human societies. Furthermore, it has been shown that
social structure is a fundamental ingredient for understanding the
distribution of sharing offers in the Ultimatum Game. We show that
selection level together with local interactions can lead to a
critical dynamics, where the precise degree of fair
offer-acceptance depends on social structure. Recent developments
on complex networks has allowed us to consider some simple models
capturing basic features of networks of interaction. Our work
emphasizes the importance of considering the social structure and
calls for the development of more realistic networks models of
social interaction to understand the interplay between individual
behavior and with whom they interact.

\section*{Acknowledgments}

We acknowledge Anxo S\'anchez, and the Editor and the referees for
providing valuable comments and insight. We acknowledge financial
support from MEC (Spain) through project FISICOS (FIS2007-60327),
and the European Commission through the NEST project EDEN
(043251); CJT acknowledges SBF (Switzerland) through SER Project
C05.0039.

\bibliographystyle{plain}

\end{document}